\documentstyle[12pt,aaspp4]{article}
\newcommand{\postscript}[2]
 {\setlength{\epsfxsize}{#2\hsize}
   \centerline{\epsfbox{#1}}}

\begin{document}

\title{Better Astrometric Deblending of Gravitational Microlensing \\
       Events by Using the Difference Image Analysis Method}

\bigskip
\bigskip
\author{Cheongho Han}
\bigskip
\affil{Department of Astronomy \& Space Science, \\
       Chungbuk National University, Chongju, Korea 361-763 \\
       cheongho@astronomy.chungbuk.ac.kr}
\authoremail{cheongho@astronomy.chungbuk.ac.kr}

\bigskip
\bigskip

\begin{abstract}
Due to the choice of very dense star fields for a higher event rate, the 
current microlensing searches suffer from large uncertainties caused by 
blending effect.  To measure light variations of microlensing events free 
from the effect of blending, a newly developed method of Differential Image 
Analysis (DIA) was proposed for microlensing searches.  However, even 
with the light variation curve obtained by using the DIA method, dramatic 
reduction of the uncertainty in the determined Einstein time scale is not
expected due to the difficulty in determining the baseline flux of a source 
star.

However, we show in this paper that if the blending effect is investigated
by detecting the shift of a source star image centroid, the DIA method will 
allow one to detect the blending effect with a significantly enhanced 
efficiency compared the efficiency of the current method based on PSF 
photometry (PSF method).  This is because for a given event the centroid 
shift measurable by using the DIA method, $\delta\theta_{\rm c,DIA}$, 
is {\it always} larger than the centroid shift measurable by using the PSF
method, $\delta\theta_{\rm c,PSF}$.  We find that the ratio 
$\delta\theta_{\rm c,DIA}/\delta\theta_{\rm c,DIA}$ rapidly increases with 
increasing fraction of blended light.  In addition, for events affected by 
the same fraction of blended light, the ratio $\delta\theta_{\rm c,DIA}/ 
\delta \theta_{\rm c,DIA}$ is larger for the event with a lower amplification. 
Therefore, centroid shift measurements by using the DIA method will be an 
efficient method to detect the blending effect especially of highly blended 
events, for which the uncertainties in the determined Einstein time scale
are large, as well as of low amplification events, for which the current 
method is highly inefficient.
\end{abstract}

\vskip20mm

\centerline{submitted to {\it the Astrophysical Journal}: Nov 10, 1999}
\centerline{Preprint: CNU-A\&SS-11/99}
\clearpage

\section{Introduction}

Searches for Galactic dark matter by detecting flux variations of source 
stars caused by gravitational microlensing have been and are being carried 
out by several groups (MACHO: Alcock et al.\ 1993; EROS: Aubourg et al.\ 
1993; OGLE: Udalski et al.\ 1993; DUO: Alard \& Guibert 1997).\markcite{
alcock93, aubourg93, udalski93, alard97}  To increase the event rate, these 
searches are being conducted towards very dense star fields such as the 
Galactic bulge and the Magellanic Clouds.  While searches towards these dense 
star fields result in an increased event rate, it also implies that the 
observed light curves are affected by the unwanted flux from unresolved 
nearby stars: blending effect.

The light curve of a microlensing event with an isolated source star is 
represented by
$$
F=A_0 F_0,
\eqno(1)
$$
where $F_0$ is the unlensed flux of the source star (baseline flux).  The 
gravitational amplification is related to the lens-source separation $u$ 
normalized by the angular Einstein ring radius by
$$
A_0 = {u^2+2 \over u\sqrt{u^2+4}};\qquad
u = \left[ {\beta_0}^2 + \left( {t-t_0\over t_{\rm E,0}}\right)^2\right]^{1/2},
\eqno(2)
$$
where the lensing parameters $\beta_0$, $t_0$, and $t_{\rm E,0}$ represent 
the lens-source impact parameter, the time of maximum amplification, and 
the Einstein ring radius crossing time scale (Einstein time scale), 
respectively.  These lensing parameters are determined by fitting theoretical 
light curves to the observed one.  One can obtain information about the lens 
mass $M$ because the Einstein time scale is proportional to the square root 
of the lens mass, i.e.\ $t_{\rm E,0}\propto M^{1/2}$.

When an event is affected by blended light, on the other hand, its light 
curve differs from that of the unblended event by
$$
F_{\rm PSF} = A_0 F_0 + B,
\eqno(3)
$$
where $B$ represents the flux from blended stars.  Then, to fit the observed 
light curve of a blended event, one should include $B$ as an additional 
parameter in addition to the three fitting parameters ($\beta_0$, $t_{0}$, 
and $t_{\rm E,0}$) of an unblended event.  As a result, the uncertainties in 
the determined Einstein time scale and the corresponding lens mass for a 
blended event are significantly larger compared to those for an unblended 
event (Di Stefano \& Esin 1995; Wo\'zniak \& Paczy\'nski 1997; Han 1997; 
Alard 1997).\markcite{distefano95, wozniak97, han97, alard97}

To resolve the blending problem, a newly developed technique to detect and 
measure light variations caused by gravitational microlensing was proposed by 
Tomaney \& Crotts (1996), Alard \& Lupton (1998), and Alard (1999).\markcite{
tomaney96, alard98, alard99}  This so-called Difference Image Analysis (DIA) 
method measures the variation of source star flux by subtracting observed 
images from a normalized reference image, i.e.\ 
$$
F_{\rm DIA} = F_{\rm obs} - F_{\rm ref} = F_0(A_0-1),
\eqno(4)
$$
where $F_{\rm obs}=A_0 F_0+B$ and $F_{\rm ref}=F_0+B$ represent the source 
star fluxes measured from the image obtained during the progress of the event 
and from the reference image, respectively.  Since not only the baseline 
flux of the lensed source star but also the flux from blended stars are 
subtracted by the DIA method, the light variation measured from the subtracted 
image is free from the effect of blending.  Since photometric precision 
is improved by removing the blended light, the DIA method was adopted by the 
MACHO group and actually applied to microlensing searches (Alcock et al.\ 
1999a, 1999b).\markcite{alcock99a, alcock99b}

However, even with the DIA method dramatic reduction of the uncertainties
in the determined Einstein time scales of gravitational microlensing events 
will be difficult.  This is because the DIA method, by its nature, has 
difficulties in measuring the baseline flux $F_0$ of a source star.  Unless 
the blended light fraction of the source star flux measured in the reference 
image, and thus the baseline flux $F_0=F_{\rm ref}-B$, is determined by some 
other means, one still has to include $B$ as an additional fitting 
parameter.\footnote{Since higher photometric precision is expected by using 
the DIA method, the uncertainties of determined lens parameters will be 
smaller than those of lens parameters determined by using the current method 
based on PSF photometry.  Han (2000) showed that for $\sim 30\%$ of high 
amplification events, one can determine $F_0$ with uncertainties less than 
50\%.}  Therefore, detecting the blending effect and estimating the blended 
light fraction in the observed source star flux is still an important issue 
to be resolved (see more discussion in \S\ 2).

There have been several methods proposed for the detection of the blending 
effect.  For a high amplification event, one can determine the unblended 
baseline flux of the source star from the shape of the light curve obtained 
by using the DIA method itself (Han 2000).  In addition, if the color 
difference between the lensed and blended stars is large, the effect of 
blending can be detected by measuring the color changes during the event 
(Buchalter, Kamionkowski, \& Rich 1996).\markcite{buchalter96}  One can also 
identify the lensed source among blended stars by using high resolution 
images obtained from the {\it Hubble Space Telescope} (HST) observations 
(Han 1997).\markcite{han97}  In addition, Han \& Kim (1999)\markcite{han99b} 
showed that the effect of blending can be detected from the astrometric 
observations of an event by using high resolution interferometers such as 
the {\it Space Interferometry Mission} (SIM).  However, these methods either 
have limited applicability only for several special cases of microlensing 
events or impractical due to the requirement of using highly-demanding 
instrument for space observations.

A much more practical method for the detection of the blending effect that 
can be applicable for general microlensing events is provided by measuring 
the linear shift of the source star image centroid towards the lensed source 
star (hereafter centroid shift) during gravitational amplification (Alard, 
Mao, \& Guibert 1995; Alard 1996; Goldberg 1998, see more detail in \S\ 
3).\markcite{alard95, alard96, goldberg98}  Goldberg \& Wo\'zniak 
(1997)\markcite{goldberg97} actually applied this method to the OGLE-1 
database and demonstrated the efficiency of this method by detecting centroid 
shifts greater than $0''\hskip-2pt .2$ for nearly half of the total 
tested events (seven out of 15 events).  However, even with this method the 
blending effect for an important fraction of blended events, especially for 
low amplification events, cannot be detected due to their small centroid 
shifts (Han, Jeong, \& Kim 1998).\markcite{han98}

In this paper, we show that if the blending effect is investigated
by detecting the shift of a source star image centroid, the DIA method will 
allow one to detect the blending effect with a significantly enhanced 
efficiency compared the efficiency of the current method based on PSF 
photometry (PSF method).  This is because for a given event the centroid 
shift measurable by using the DIA method, $\delta\theta_{\rm c,DIA}$, 
is {\it always} larger than the centroid shift measurable by using the PSF 
method, $\delta\theta_{\rm c,PSF}$.  We find that the ratio 
$\delta\theta_{\rm c,DIA}/\delta\theta_{\rm c,DIA}$ rapidly increases with 
increasing fraction of blended light.  In addition, for events affected by 
the same fraction of blended light, the ratio $\delta\theta_{\rm c,DIA}/ 
\delta \theta_{\rm c,DIA}$ is larger for the event with a lower amplification.  
Therefore, centroid shift measurements by using the DIA method will be an 
efficient method to detect the blending effect especially of highly blended 
events, for which the uncertainties in the determined Einstein time scale are
large, as well as of low amplification events, for which the current method 
is highly inefficient.

\section{Degeneracy Problem}

Even with the blending-free flux variations of a gravitational microlensing 
event measured by using the DIA method, it will be difficult to know whether 
the event is affected by the blending effect or not.  This because the 
best-fit light curve obtained under the wrong assumption that the event is 
not affected by the blending effect matches well with the observed light 
curve.

The relations between the best-fit lensing parameters of a microlensing event 
resulting from the wrong determination of its source star baseline flux
and their corresponding true values are provided by the analytic equations 
derived by Han (1999).\markcite{han99a}  If a blended event is misunderstood 
as an unblended event, the baseline flux of the source star is overestimated 
by $F_0+B$, causing mis-normalization of the amplification curve.\footnote{
The term `amplification curve' represents the changes in the amplification 
of the source star flux as a function of time.}  Then the best-fit impact 
parameter $\beta$ determined from the mis-normalized amplification curve 
differs from the true value $\beta_0$ by
$$
\beta = \left[ 2(1-A_{\rm p}^{-2})^{-1/2}-2\right]^{1/2};\qquad
A_{\rm p} = {A_{\rm p,0}+\eta\over 1+\eta},
\eqno(5)
$$
where $A_{\rm p,0}= (\beta_0^2+2)/(\beta_0\sqrt{\beta_0^2+4})$ and $A_{\rm p}$
represent the peak amplifications of the true and the mis-normalized 
amplification curves and $\eta=\Delta F_0/F_0=B/F_0$ is the fractional 
deviation of the mis-determined baseline flux.  Mis-normalization of the 
amplification curve makes the best-fit Einstein time scale also differ
from the true value by
$$
t_{\rm E}=t_{\rm E,0}\left( {{\beta_{\rm th}}^2-{\beta_0}^2 
\over 
{\beta_{\rm th,0}}^2-{\beta}^2} \right)^{1/2}.
\eqno(6)
$$
Here $\beta_{\rm th,0}=1.0$ and $A_{\rm th,0}=3/\sqrt{5}$ represent the 
threshold impact parameter and the corresponding threshold amplification for 
event detection, respectively.  However, due to the mis-normalization of the 
amplification curve, the actually applied threshold amplification and the 
corresponding threshold impact parameter differ from $A_{\rm th,0}$ and 
$\beta_{\rm th,0}$ by
$$
A_{\rm th}=A_{\rm th,0} (1+\eta)-\eta
\eqno(7)
$$
and
$$
\beta_{\rm th}=\left[ 2(1-A_{\rm th}^{-2})^{-1/2}-2 
\right]^{1/2}.
\eqno(8)
$$

Figure 1 shows the degeneracy problem in the light curve measured by using 
the DIA method.  In the figure, the solid curve presents the light variation 
curve  when an example event with lensing parameters $\beta_0=0.5$ and 
$t_{\rm E,0}=1.0$ is observed by using the DIA method.  The event has baseline 
flux of $F_0=0.5$ and it is affected by the blended flux of $B=0.5$.  The 
dotted curve represents the best-fit light curve obtained by assuming that 
the event is not affected by the blending effect.  The best-fit lensing 
parameters of the mis-normalized light curve determined by using the relations 
in equations (5)--(8) are $\beta=0.756$ and $t_{\rm E}=0.742$, respectively.  
One finds that the two light curve matches very well, implying that it will 
be not easy to detect the blending effect only from the light variation curve 
obtained by using the DIA method.

\section{Centroid Shifts by Using the PSF and DIA Methods}

In previous section, we show that detection of the blending effect for 
general microlensing events will be difficult even with the blending-free 
light variation curve obtained by using the DIA method.  However, we show 
in this section that if the blending effect of a microlensing event is 
investigated by detecting the centroid shift of a source star image, the 
DIA method will allow one to detect the blending effect with a significantly 
increased efficiency than the current method based on PSF photometry does.  
This is because for a given event the centroid shift measurable by using 
the DIA method is {\it always} larger than the shift measurable by the PSF 
method.

For the comparison of the centroid shifts of an event measurable by the DIA 
and the PSF methods, let us consider a blended source star image within which 
multiple stars with individual positions and fluxes ${\bf x}_i$ and $F_{0,i}$ 
are included.  Gravitational amplification occurs only for one of these 
stars.\footnote{For Galactic microlensing event, the typical lens-source 
separation is of the order of milli-arcsec, while the average separation 
between stars is $({\cal O}) 10^{-1}$ arcsec.  Therefore, simultaneous 
lensing for multiple source stars hardly happens.} Since the position of 
the lensed star usually differs from the centroid of the blended image, the 
centroid of the source star image is shifted toward the lensed source during 
the event.  If the lensed source star with a baseline flux $F_{0,j}$ is 
located at a position ${\bf x}_{j}$, the amount of this centroid shift, 
which is measurable by using the current PSF method, is
$$
\vec{\delta\theta}_{\rm c,PSF} = {\cal D} 
\left( \langle{\bf x}\rangle -{\bf x}_{j}\right) ;\qquad
{\cal D} = {f(A_0-1) \over f(A_0-1)+1},
\eqno(9)
$$
where $f=F_{0,j}/\sum_{i}F_{0,i}$ represents the fractional flux of the 
lensed source out of the total flux of the blended stars (including $F_{0,j}$)
within the integrated seeing disk and $\langle{\bf x}\rangle=\sum_i {\bf x}_{i} 
F_{0,i}/\sum_i F_{0,i}$ is the position of blended image centroid before 
the gravitational amplification, i.e.\ the position of the blended source 
image centroid on the reference frame (Goldberg 1998).\markcite{goldberg98}  
On the other hand, {\it the position of the centroid measured on the 
subtracted image by using the DIA method is the true position of the lensed 
source star}, i.e.\ ${\bf x}_{j}$.  Therefore, the centroid shift measureable 
by using the DIA method is 
$$
\vec{\delta\theta}_{\rm c,PSF} = 
\langle{\bf x}\rangle -{\bf x}_{j} .
\eqno(10)
$$
Then the ratio between the centroid shifts measurable by the two methods
is $\delta\theta_{\rm c,DIA}/\delta\theta_{\rm c,PSF}={\cal D}^{-1}$.  Since 
$f(A_0-1)>0$, and thus ${\cal D}<1$, for a given event the centroid shift 
measurable by the DIA method is always larger than the shift measurable by 
the PSF method.

In Figure 2, we illustrate the centroid shifts measurable by the DIA and the 
PSF methods for visualization.  On the left side, we present the contours 
(measured at an arbitrary flux level) of two source stars with identical 
baseline fluxes (the inner two dotted circles with their centers marked by 
`+') and their integrated images (the outer solid curve centered at `x') 
before (upper part) and during (lower part) gravitational amplification.  
Among the two stars within the blended image, the right one is lensed. 
Between the two left contours of integrated image, we marked the centroid 
shifts measurable by the PSF ($\vec{\delta\theta}_{\rm c,PSF}$) and the DIA 
($\vec{\delta\theta}_{\rm c,DIA}$) methods.  To better show the centroids 
shifts, the region enclosed by a dot-dashed line is expanded and presented on 
the right side.  One finds that 
$\delta\theta_{\rm c,DIA}>\delta\theta_{\rm c,PSF}$.

Then, how much larger is the centroid shift measurable by the DIA method than 
the shift measurable by the DIA method?  The ratio  $\delta\theta_{\rm c,DIA}/
\delta\theta_{\rm c,PSF}$ depends not only on the blended light fraction 
$1-f$, but also on the amplification $A_0$ of an event.  To see these 
dependencies, we present in Figure 3 the relations between the ratio 
$\delta\theta_{\rm c,DIA}/ \delta\theta_{\rm c,PSF}$ and the fraction of 
blended light for events with various impact parameters.  
From the figure, one finds the two following imporant trends.  First, the 
ratio $\delta\theta_{\rm c,DIA}/\delta\theta_{\rm c,PSF}$ increases rapidly 
with the increasing fraction of blended light.  This implies that compared to 
the PSF method the DIA method will be able to better detect the blending 
effect of highly blended events for which the uncertainties in the determined 
Einstein time scales are large.  Second, if events are affected by the same 
fraction of blended light, the ratio $\delta\theta_{\rm c,DIA}/\delta
\theta_{\rm c,PSF}$ is bigger for the event with lower amplification, implying 
that the blending effect of low amplification events can be better detected 
by using the DIA method.  For events with low amplifications, the expected 
centroid shifts measurable by the PSF method are very small, making it 
difficult to detect the blending effect by this method.  Therefore, by 
increasing the detection efficiency for centroid shifts for low amplification 
events, the DIA method will allow one to detect the blending effect for a 
significant fraction of Galactic microlensing events.

\section{Summary}

We analyze the centroid shifts of the source star image centroid of 
gravitational microlensing events caused by the blending effect.  The findings 
from the comparison of the centroid shifts measurable by the current method 
based on PSF photometry and the newly developed DIA method are summarized as 
follows.

\begin{enumerate}
\item
For a given event the centroid shift measurable by using the DIA method is 
always larger than the shift measurable by using the PSF method, allowing 
one to better detect the blending effect of microlenisng events.

\item
The ratio between the centroid shifts measurable by using the DIA and the 
PSF methods, $\delta\theta_{\rm c,DIA}/\delta\theta_{\rm c,DIA}$, rapidly 
increases as the fraction of the blended light increases.  Therefore, 
detection of the centroid shifts by using the DIA method is an efficient 
method for the detection of the blending effect especially for highly blended 
events which cause large uncertainties in the determined Einstein time scales.

\item
If the blended light fraction is the same, the ratio $\delta\theta_{\rm c,DIA}/
\delta\theta_{\rm c,DIA}$ is larger for the events with a lower amplification.
Therefore, the DIA method enables one to better detect the source star image 
centroid shifts of low amplification events, for which 
detection of their blending effect by using the PSF is very difficult
due to their small expected shifts.

\end{enumerate}

\acknowledgements
This work was supported by the grant (1999-2-113-001-5) from Korea Science 
\& Engineering Foundation (KOSEF).

\clearpage

\postscript{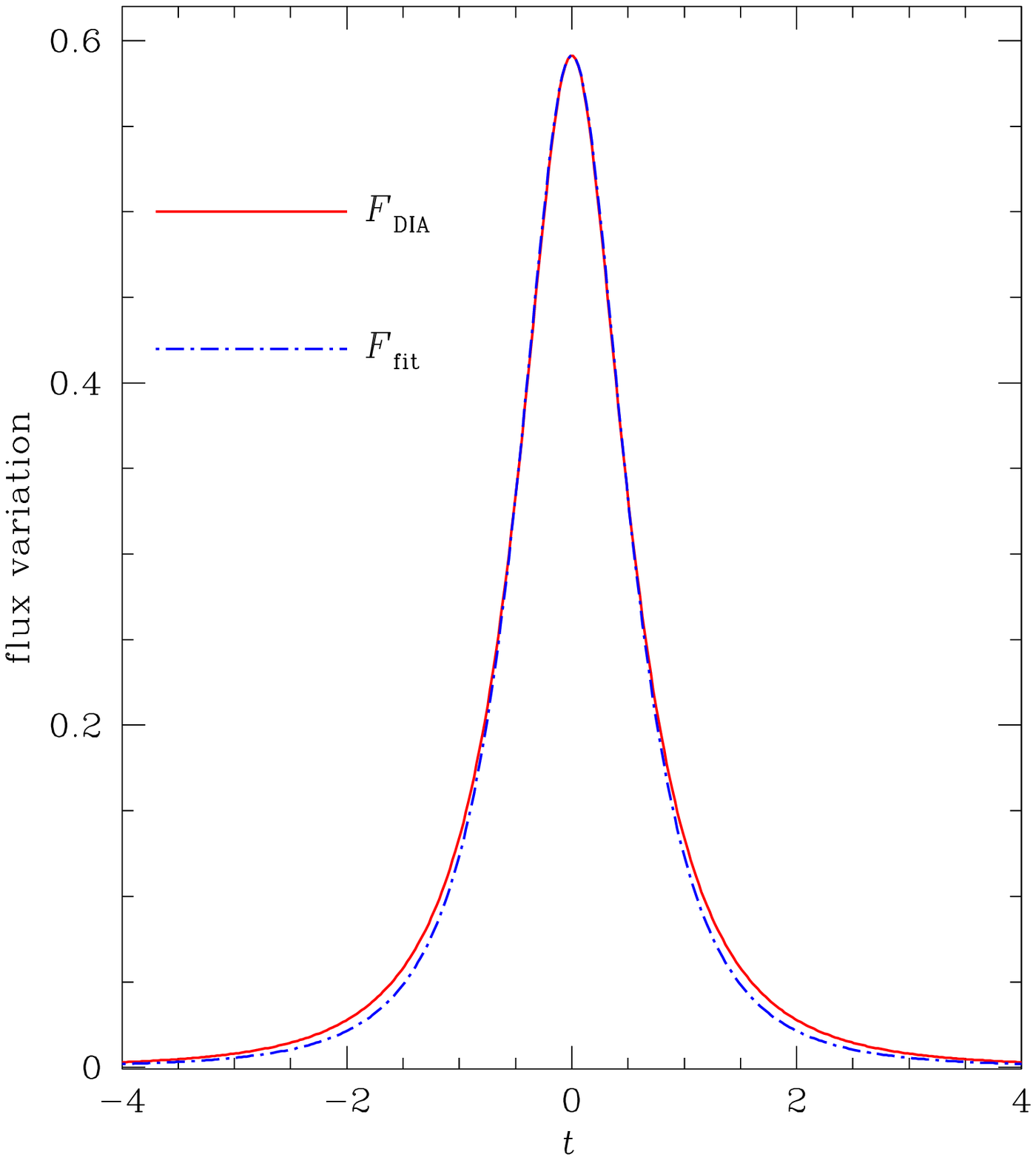}{1.0}
\noindent
{\footnotesize {\bf Figure 1:}\ 
The degeneracy problem in the light variation curve of a gravitational 
microlensing event observed by the DIA method.  The solid curve represents 
the light variation curve which is expected when an example event with 
$\beta_0=0.5$, $t_{\rm E,0}=1.0$, and $F_0=0.5$ is observed by using the 
DIA method.  The dot-dashed curve represents the best-fit curve obtained 
under the assumption that the flux of the source star is not affected by 
the blending effect, despite that the source star flux in the reference 
image is influenced by the blended light with an amount $B=0.5$.
}\clearpage

\postscript{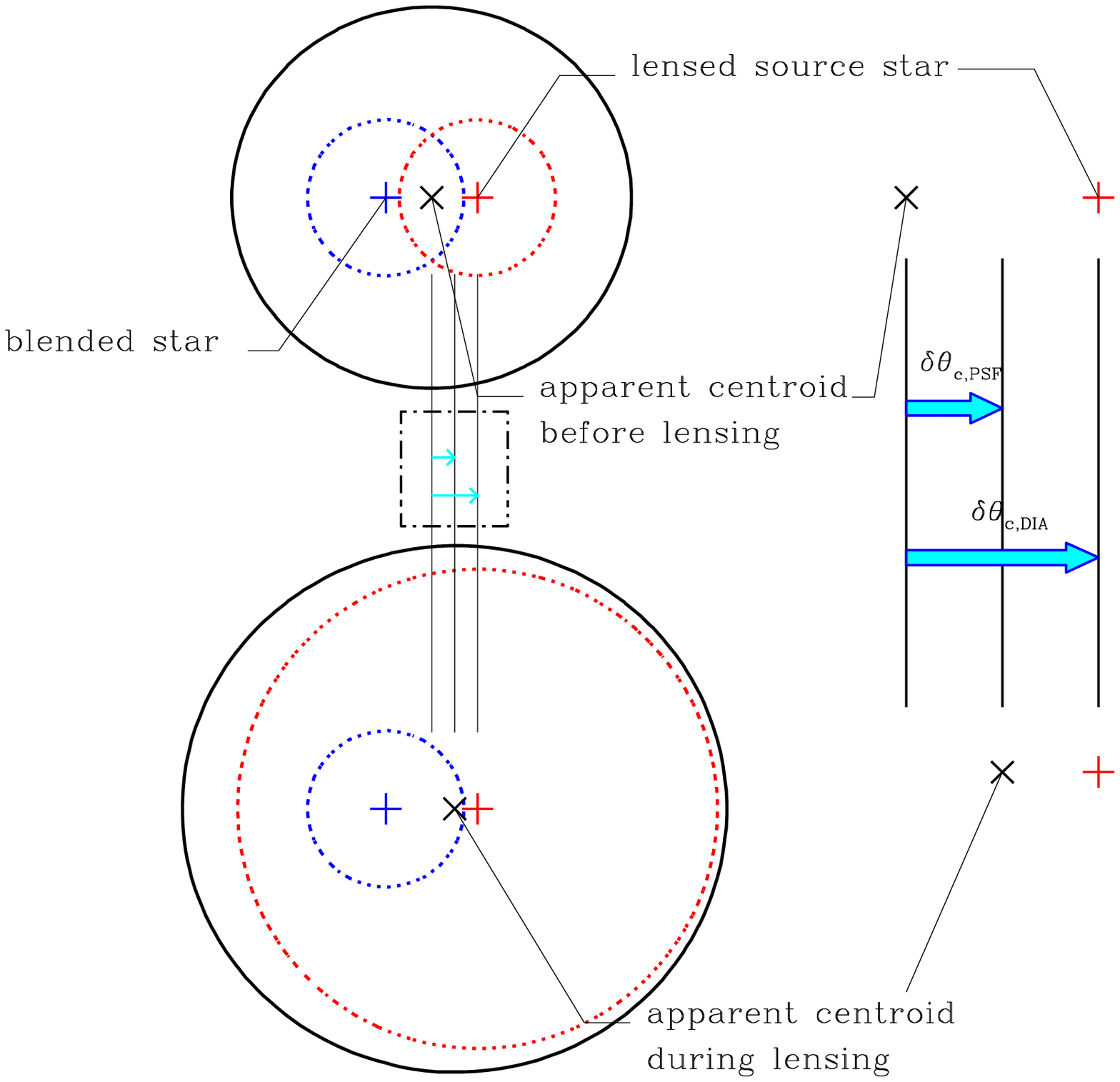}{1.0}
\noindent
{\footnotesize {\bf Figure 2:}\ 
Illustration of the source star image centroid shifts measurable by usng the 
PSF ($\delta\theta_{\rm c,PSF}$) and the DIA ($\delta\theta_{\rm c,DIA}$) 
methods.  On the left side, the contours (measured at an arbitrary flux 
level) of two source stars with identical baseline fluxes (the inner two 
dotted circles with their centers marked by `+') and their integrated images 
(the outer solid curve centered at `x') before (upper part) and during (lower 
part) gravitational amplification are presented.  Among the two stars within 
the blended image, the right one is lensed.  Between the two left contours 
of integrated image, the centroid shifts measurable by the PSF ($\vec{\delta
\theta}_{\rm c,PSF}$) and the DIA ($\vec{\delta\theta}_{\rm c,DIA}$) methods 
are marked.  To better show the centroids shifts, the region enclosed by a 
dot-dashed line is expanded and presented on the right side.  
}\clearpage

\postscript{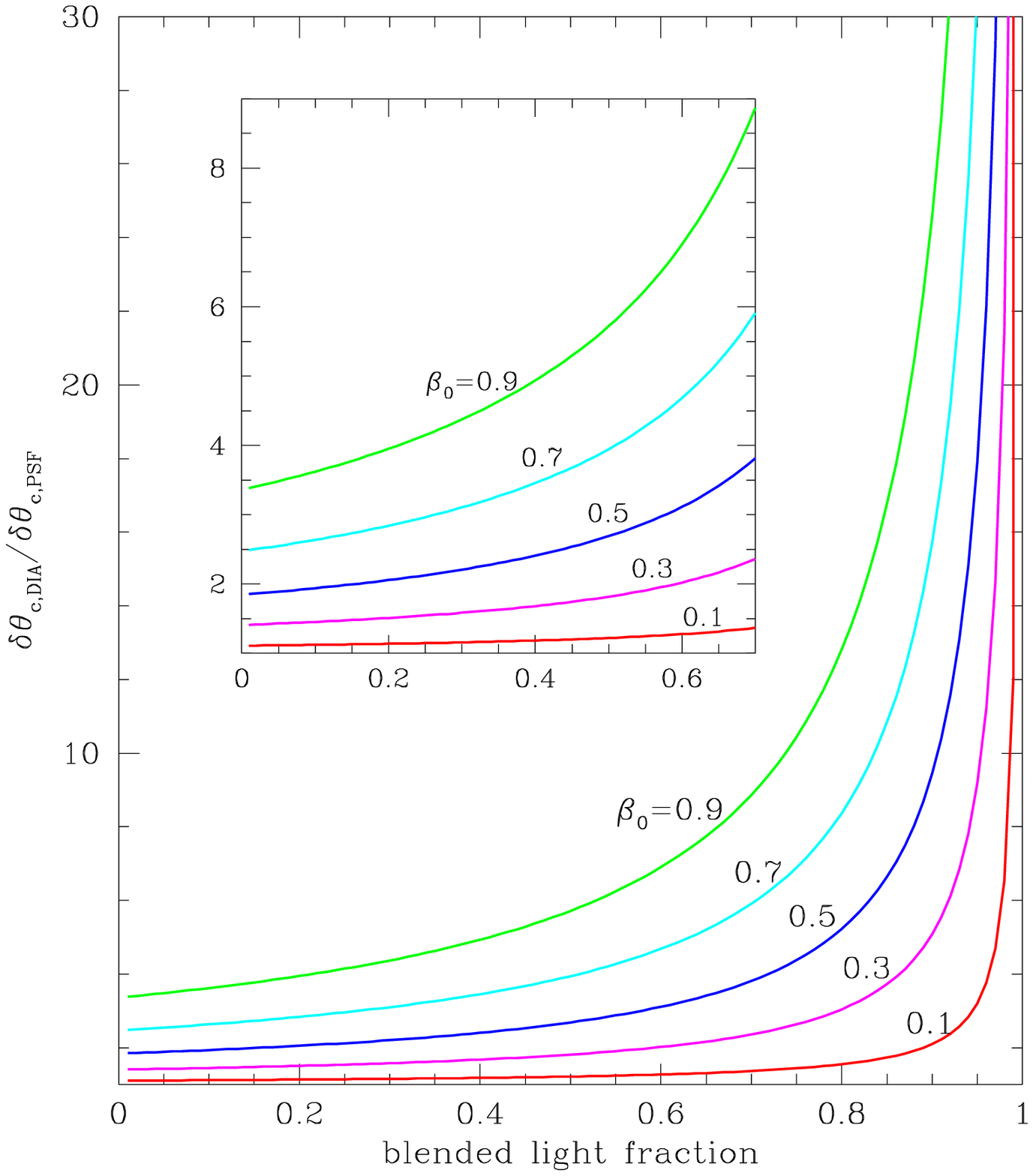}{1.0}
\noindent
{\footnotesize {\bf Figure 3:}\ 
The relations between the centroid shift ratio $\delta\theta_{\rm c,DIA}/
\delta\theta_{\rm c,PSF}$ and the fraction of blended light $1-f$ for events 
with various impact parameters.  To better show the relations with 
$(1-f)\leq 0.7$, the region is enlarged and presented in a separate box.  
}\clearpage

\end{document}